\documentclass[superscriptaddress,nofootinbib, amsmath,amssymb,preprintnumbers,prdfloatfix,twocolumn,PRD]{revtex4-2}
\usepackage{CJK}
\usepackage{graphicx}% Include figure files
\usepackage{dcolumn}% Align table columns on decimal point
\usepackage{upgreek}
\usepackage{amsmath}
\usepackage{bm}% bold math
\usepackage[colorlinks=true,pdfstartview=FitV,breaklinks=true]{hyperref}
\usepackage[dvipsnames,table]{xcolor}
\hypersetup{urlcolor=BlueViolet,
	    citecolor=Plum,
	    linkcolor=PineGreen}
\usepackage{lipsum}	    
\usepackage{titlesec}
\usepackage{etoolbox}% http://ctan.org/pkg/etoolbox
\usepackage[normalem]{ulem}
\newcommand{\fdm}{f_\mathrm{DM}}

\newcommand{\dndt}{\frac{\mathrm{d}n}{\mathrm{d}t}}
\newcommand{\rhodm}{\rho_{\mathrm{DM}}}
\newcommand{\fMyr}{f_{\mathrm{Myr}}}
\definecolor{offblue}{RGB}{23,80,153}

\usepackage[export]{adjustbox}

\begin{document}

\preprint{IPMU23-0017}

\title{Constraints on late-forming exploding black holes}

 \author{Zachary S. C. Picker}
 \email{zpicker@physics.ucla.edu}

\affiliation{Department of Physics and Astronomy, University of California Los Angeles,\\ Los Angeles, California, 90095-1547, USA}

\author{Alexander Kusenko}
\email{kusenko@ucla.edu}
    \affiliation{Department of Physics and Astronomy, University of California Los Angeles,\\ Los Angeles, California, 90095-1547, USA}
    \affiliation{Kavli Institute for the Physics and Mathematics of the Universe (WPI), The University of Tokyo Institutes for Advanced Study, The University of Tokyo, Chiba 277-8583, Japan}
    
\begin{abstract}
\noindent 
Black holes can be produced in collapse of small-scale dark matter structures, which can happen at any time from the early to present-day universe.   Microstructure black holes (MSBHs) can have a wide range of masses. Small MSBHs evaporate via Hawking radiation with lifetimes shorter than the age of the universe, but they are not subject to the usual early-universe bounds on the abundance of small primordial black holes. We investigate the possible signal of such a population of exploding, late-forming black holes, constraining their abundance with observations from diffuse extragalactic gamma- and x-ray sources, the galactic center, and dwarf spheroidal galaxies. 
\end{abstract}

\maketitle

\section{INTRODUCTION} 
\noindent Primordial black holes (PBHs)~\cite{pbh,Hawking:1971ei,Carr:1974nx,Chapline:1975ojl}, which form from non-stellar processes, are generally understood to form in the very early universe~\cite{carr_constraints_2021,green_primordial_2020}. One of the primary ways to constrain the abundances of the least massive PBHs is from Hawking radiation~\cite{Hawking:1974rv,Hawking:1974sw}---black holes are understood to emit a nearly-blackbody spectrum, with temperature proportional to the surface gravity at their horizon. Evaporation from Hawking radiation means that black holes smaller than the `critical mass', $m\sim 8\times10^{14}~g$, have lifetimes shorter than the age of the universe~\cite{MacGibbon:2007yq,arbey_blackhawk_2019}. Because of this, existing constraints on the abundance of these small evaporating black holes are exclusively placed by considering their effect on early-universe processes, such as the cosmic microwave background anisotropies or big bang nucleosynthesis~\cite{carr_new_2010,Acharya:2020jbv,Chluba:2020oip,carr_constraints_2021,green_primordial_2020}.

However, there has been increasing interest in `nonstandard' black hole formation mechanisms---in particular, there is the possibility that black holes could form from the collapse of dark matter structures, such as halos produced by Yukawa forces, or Q-balls, or Fermi balls~\cite{flores_primordial_2021,chakraborty_formation_2022,lu_late-forming_2022,kawana_primordial_2022,Domenech:2023afs}. If the timescale of collapse is slow for these structures, it is possible to produce black holes extremely late in the universe, even into the future~\cite{lu_late-forming_2022,Domenech:2023afs}. We call these black holes microstructure black holes (MSBH) in order to distinguish them from both PBH and stellar black holes (which naturally form in late times). In contrast with stellar black holes, MSBHs could form over a large range of masses, depending on the dark matter structures they form from. If they are much smaller than the critical mass, they will `explode' over short timescales compared to the age of the universe, producing large fluxes of photons and other particles. In the most realistic scenario, where the dark matter structures themselves have some distribution in masses, we would therefore have a continuous `injection' of these exploding black holes. Importantly, this exploding population would evade all the early universe bounds which are usually used to constraint PBHs at these masses.

\begin{figure}[!ht]
    \centering
    \includegraphics[width=\columnwidth]{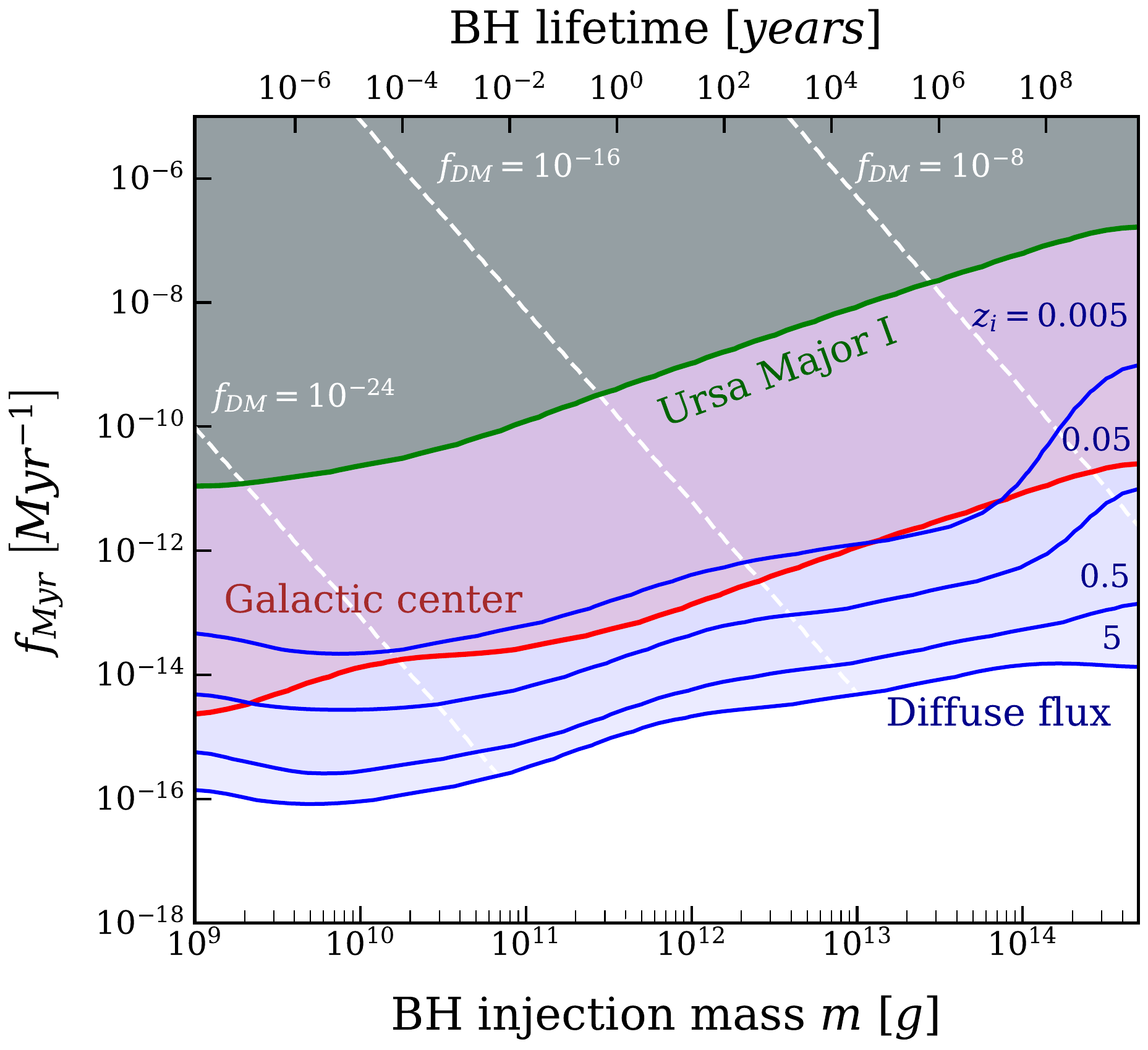}
\caption{Our final constraints for microstructure black holes, as a function of injection mass $m$ and injection fraction per one million years $\fMyr$. The red region corresponds to constraints placed from the galactic center. The blue region represents constraints from the diffuse extragalactic gamma-ray and x-ray flux, where the four different blue bands correspond to different ranges of the injection time between redshifts $z_f=0$ and $z_i=0.005-5$. The green region corresponds to the dwarf spheroidal constraints---specifically, the largest signal would be from Ursa Major I. The dashed white lines give the mass density of the MSBH population as a fraction of the total dark matter density from Eq.~\ref{eq:bhdm}.}\label{fig:final_constraint}
\end{figure}

In this paper we will investigate the constraints on such an exploding population of late-forming black holes from a phenomenological viewpoint. In Sec.~\ref{sec:model} we outline our model and derive a generic mass distribution for injected, evaporating black holes. In Sec.~\ref{sec:diffuse} we construct constraints on this population based on the diffuse extragalactic gamma-ray and x-ray flux, and in Sec.~\ref{sec:center} we examine the galactic center gamma-ray flux. In Sec.~\ref{sec:dwarf} we add the dwarf spheroidal constraints and finally estimate the rate of nearby explosions in Sec.~\ref{sec:explode} before concluding.

\section{MODELING THE MSBH POPULATION}\label{sec:model}
\noindent Taking inspiration from scenarios such as those in  Refs.~\cite{lu_late-forming_2022,flores_inhomogeneous_2022,flores_primordial_2021,savastano_primordial_2019,hong_fermi-ball_2020,lu_old_2022,kawana_first-order_2022,kawana_primordial_2022,Domenech:2023afs}, let us assume that we have some distribution of dark structures, which are slowly collapsing as they lose energy (e.g., from scalar radiation and cooling). Once these structures reach a particular threshold condition, they completely collapse, forming a black hole at a fixed mass related to the physics of the collapse. We can then consider MSBH formation a kind of black hole `injection' which traces the dark matter distribution. We will consider here black holes smaller than about $8\times 10^{14}~g$, since that is the black hole which has a lifetime roughly the age of the universe. For larger black holes, the usual PBH abundance constraints from observations today---such as the galactic center---will apply straightforwardly, with the overall MSBH abundance given approximately by the product of the injection period and injection rate. We also cut off our mass range arbitrarily at $10^9~g$, since they have a lifetime shorter than a second.

If the dark matter structures are formed from anisotropies and mergers in the early universe, they will most likely have some extended mass distribution. For the sake of simplicity, we will assume this is a flat distribution, so that MSBHs are injected at a constant rate between the time when the structures first start collapsing, and when they have all collapsed (but we note that a more complicated dark structure distribution might be interesting to study, since it would lead to a time-varying MSBH injection rate). We technically remain agnostic as to the fraction of dark matter in these structures, but a higher abundance of dark matter structures would naturally lead to higher injection rates. We then define the black hole injection rate generically as,
\begin{align}
    \dndt(\mathbf{x}) = \frac{\fMyr}{m}~ \rho_{DM}(\mathbf{x})~,
\end{align}
where $m$ is the central mass of the black holes being formed, $\rhodm$ is the dark matter energy density, and the important quantity $\fMyr$ is defined to be the fraction of dark matter which is converted to black holes in a Myr period. These quantities are allowed to be spacetime-dependent, since the dark matter density itself varies with time and position within the galaxy. We also need to parametrize the start time of MSBH formation at redshift $z_i$ and end time at redshift $z_f$. These two times are of course related to the initial dark matter structure distribution, since we might presume that the minimal size structures will collapse first and the maximal sized structures last. 

We should require that the dark matter energy density not change significantly over the period of black hole injection. The most conservative assumption would be to assume that the MSBHs are injected over the lifetime of the universe, and to set an arbitrary cutoff that $0.1\%$ of the dark matter at most could be lost in this time period. This is probably a conservative estimate since current observations of the Milky Way's mass loss are only capable of constraining the total loss to about 2\% per Gyr~\cite{Sharma:2022qtw}. We then have an upper bound given by,
\begin{align}
    (1-\fMyr~Myr)^{t/Myr}~\rho_{\rm DM}&\lesssim 0.001 \rho_{\rm DM}\nonumber\\
    \fMyr&\lesssim 5\times10^{-4}~\mathrm{Myr}^{-1}~,
\end{align}
where $t$ is here the time in Myrs.

\subsection{The triangle distribution}

\noindent The continuous injection of black holes at mass $m$ will lead to a \textit{static} extended mass distribution for the black holes over a time period which is longer than the black hole lifetime, and where the change in injection rate is negligible. Because this distribution is static, we can define the fraction of total black holes in the range $[M,M+\mathrm{d}M]$ as,
\begin{equation}\label{eq:phi}
    \phi(M) \equiv \frac{1}{n_\mathrm{BH}}\frac{\mathrm{d}n(M)}{\mathrm{d}M}~,
\end{equation}
where $n_\mathrm{BH} = \dndt \tau(m)$ is the total BH number density and $n(M)$ is the number density of BHs in the mass range $[M,M+\mathrm{d}M]$. If the injection mass was \textit{exactly} monochromatic, this distribution would be simply flat---however, it is much more realistic to assume that there would at least some extension to the mass distribution of the MSBHs. In particular, it is known~\cite{carr_constraints_2016,Cai:2021fgm,mosbech_effects_2022} that evaporating black holes in any sufficiently well-behaved non-monochromatic mass distribution evolve into a characteristic `triangle' distribution. This unique distribution arises because of the nonlinear nature of evaporation---a black hole at mass $M$ evolves at a different rate to the black hole at mass $M+\mathrm{d}M$. As a result, the small section of distribution just above $M$ is `stretched' into a triangle-shaped wedge (in log-log space), with slope proportional to $M^2$. This slope is due to the mass-loss equation for black holes~\cite{page1976ApJ...206....1P}:
\begin{align}\label{eq:bhdm}
    \frac{dM}{dt} = -\frac{\hbar c^4}{G^2}\frac{\alpha_\mathrm{eff}(M)}{M^2},
\end{align}
where following Ref.~\cite{mosbech_effects_2022} we have defined an effective parameter $\alpha_\mathrm{eff}(M)$ to account for the changes in permissible particle species that the black hole can emit at very small masses\footnote{$\alpha_\mathrm{eff}=1/15360\pi$ for large black holes emitting only photons. For small black holes, we must calculate this parameter numerically, for example with the code BlackHawk~\cite{arbey_blackhawk_2019,arbey_physics_2021,mosbech_effects_2022}. For reference, $\alpha_\mathrm{eff}(10^{18}~g)\sim7\times10^{-4}$ and $\alpha_\mathrm{eff}(10^{10}~g)\sim3\times10^{-3}$.}. Since $\alpha_\mathrm{eff}$ doesn't vary by more than an order of magnitude for evaporating black holes~\cite{mosbech_effects_2022}, for non-monochromatic initial mass distributions, the shape of the evolved triangular tail is dominated by the $M^{-2}$ term in Eq.~\ref{eq:bhdm}. We ignore the modification of $\alpha_\mathrm{eff}$ due to the additional emission of extra dark matter species, since this which would not significantly modify our results. 

\begin{figure}[!ht]
    \centering
    \includegraphics[width=\columnwidth]{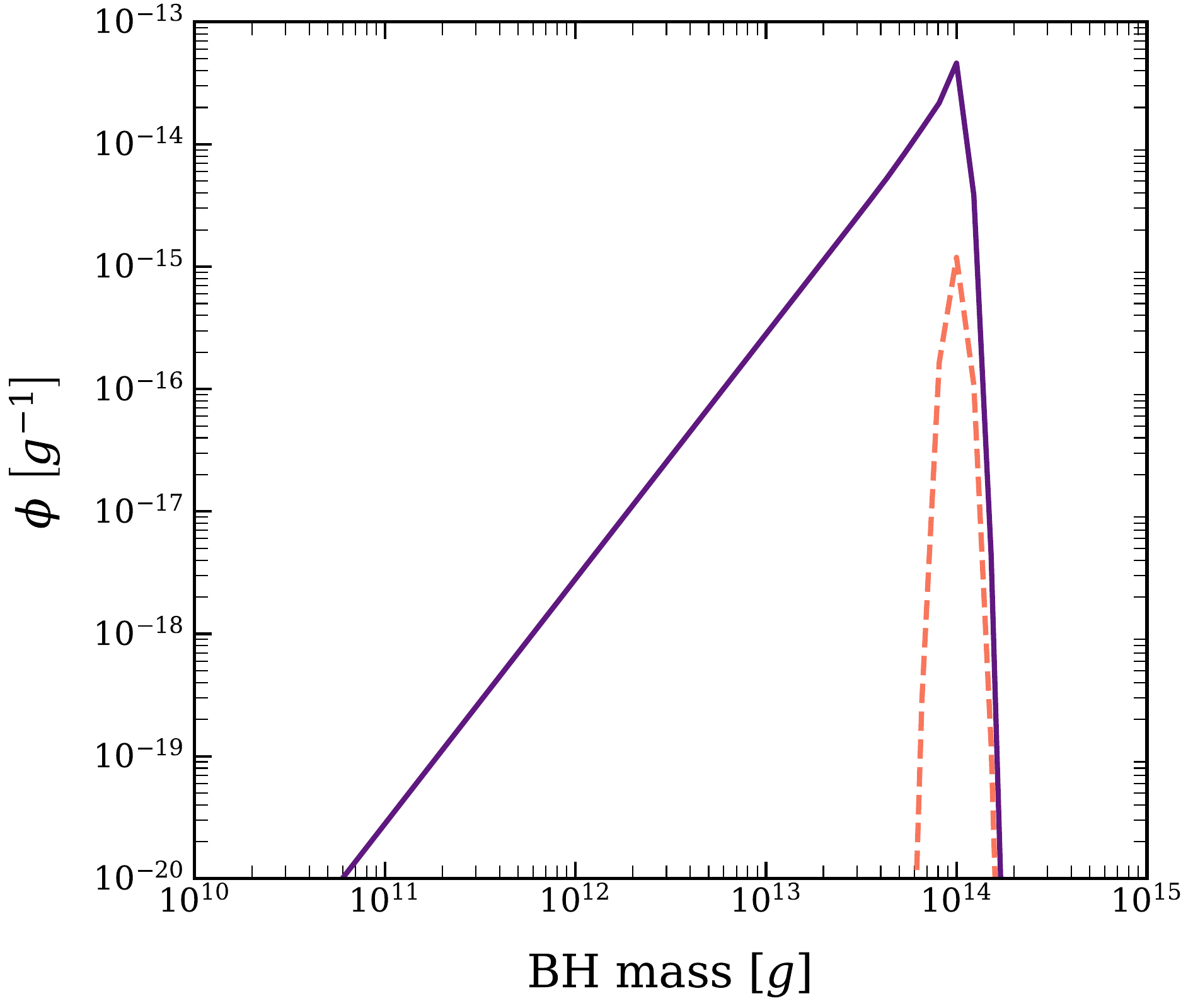}
    \caption{The solid purple curve shows the extended distribution for $m=10^{14}~g$ and deviation $\sigma=0.1$, integrated over a Gyr period (note that the black hole lifetime $\tau(10^{14}~g)\sim30~Myr$). The dashed orange curve shows the distribution of of black holes injected in a single Myr period.}\label{fig:distrnum}
\end{figure}

Our injected triangle distribution is shown explicitly in Fig.~\ref{fig:distrnum}, where we have numerically simulated the injection of narrow ($\sigma=0.1$) lognormally-distributed MSBHs centered at $m=10^{14}~g$ over a sufficiently long period of time.~\footnote{The lognormal distribution is given by,
\begin{align}
    \frac{\mathrm{d}n(M)}{\mathrm{d}M} = \frac{n_\mathrm{BH}}{\sqrt{2\pi}\sigma M} \exp{\left( -\frac{(\ln (M/m))^2}{2\sigma^2} \right)}~,
\end{align}
but the shape of this distribution is not important to our results at all as long as the actual distribution is sufficiently narrow.}

Since our continually-injected distribution is static, we can use that
\begin{align}
1 = \int_0^m \mathrm{d}M\phi(M)~,
\end{align}
and the fact that the slope of the low-mass tail is proportional to $M^2$ to find the maximum fraction as $\phi(m) = 3/m$.  In the numerical simulation in Fig.~\ref{fig:distrnum}, $\phi(m)\sim2.2/m$ as a result of the broadness of the lognormally-distributed black holes. The narrower the distribution, however, the closer we get to the theoretical value. Because $\phi(M)$ is defined fractionally, the plot in Fig.~\ref{fig:distrnum} is independent of the injection rate.

Finally, it might be useful to express our population of exploding black holes as a fraction of the total dark matter energy density, similar to the usual way where $\rho_{\mathrm{BH}}\equiv \fdm\rhodm$. We can use that,
\begin{align}
    \rho_\mathrm{BH} \equiv n_\mathrm{BH}\int_0^m M \phi(M) \mathrm{d}M~.
\end{align}
and that the slope of our distribution function goes like $\beta M^2$, with $\phi(m)=3/m$, so that $\beta=3/m^3$. Then we have,
\begin{align}
    \fdm\rhodm &= \dndt \tau(m)\int_0^m 3\frac{M^3}{m^3}\mathrm{d}M\nonumber\\
    \fdm &= \frac34\tau(m)\fMyr~,
\end{align}
where $\tau(m)$ is again the lifetime expressed in Myr. Indeed, when the lifetime of the black holes is smaller than a Myr, the fraction of dark matter in these black holes is smaller than $\fMyr$, as we might expect (if the lifetime was exactly a Myr, we would still have a smaller fraction, since the black holes are losing mass to radiation as they evaporate).

\section{EXTRAGALACTIC GAMMA-RAY CONSTRAINTS}\label{sec:diffuse}
\begin{figure}[!ht]
    \centering
    \includegraphics[width=\columnwidth,valign=t]{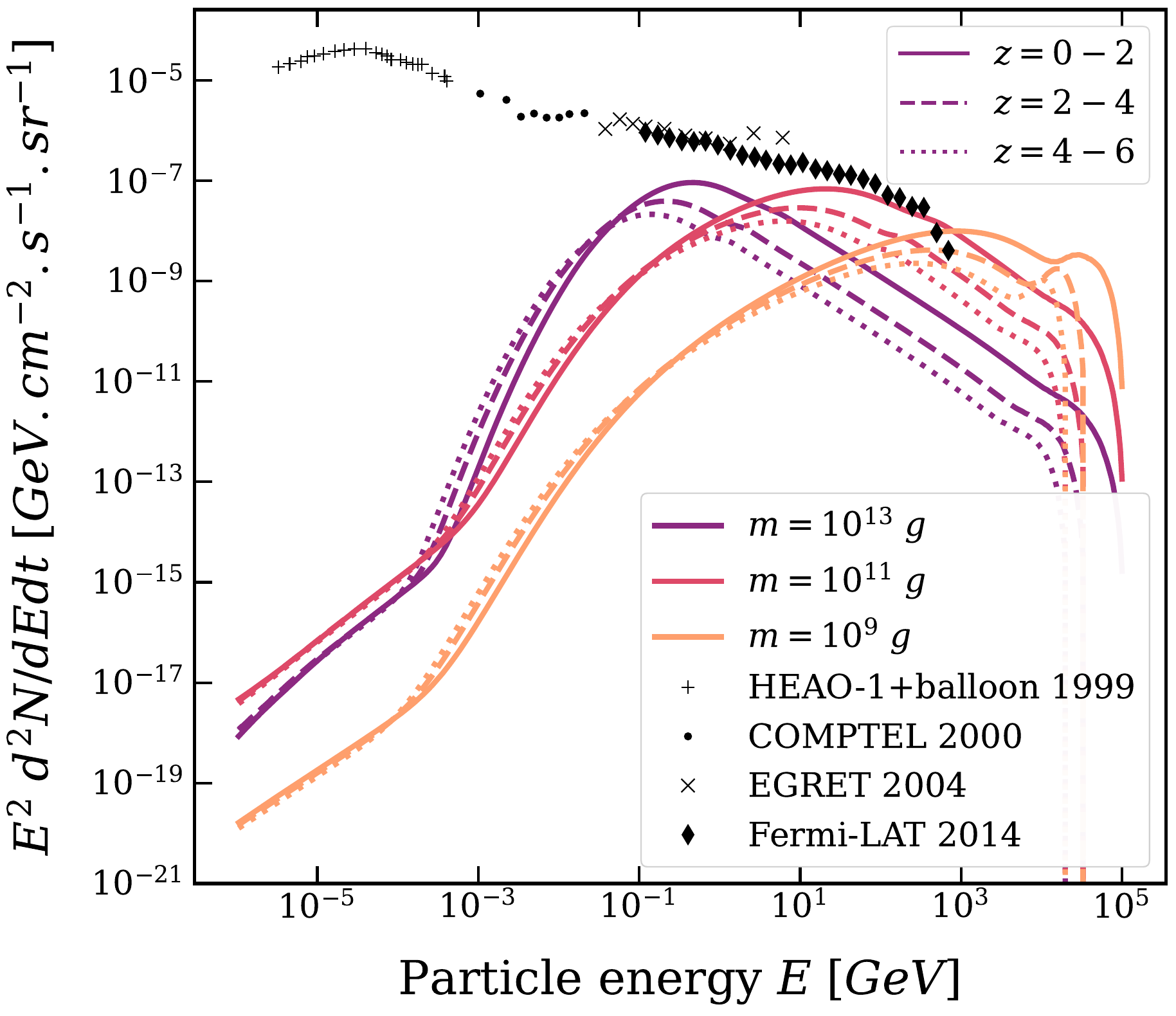}
\caption{Curves show the extragalactic gamma-ray differential flux for three masses of injected black hole, and for three choices each of injection timeframe. For the sake of comparison, the observations are similar to those given in Ref.~\cite{carr_new_2010}: COMPTEL~\cite{2000thesymposium,carr_new_2010}, EGRET~\cite{Strong:2004ry} and HEAO-1 and other balloon observations~\cite{Gruber:1999yr} are plotted, as well as the FERMI-LAT~\cite{Fermi-LAT:2014ryh} measurements of the isotropic gamma-ray background. Here we have set $\fMyr=10^{-15}$.}\label{fig:diffuse_photons}
\end{figure}

\noindent The explosion of MSBHs will lead to a diffuse extragalactic gamma ray signal, in the same manner as was previously explored for PBHs in e.g. Ref.~\cite{carr_new_2010}. The particle spectra here and throughout was found using the program BlackHawk~\cite{arbey_blackhawk_2019,arbey_physics_2021,Sjostrand:2014zea,Bellm:2015jjp,Coogan:2019qpu,Bauer:2020jay}. Although the universe is of course inhomogenous and dark matter is found in halos and large scale structures, we can more simply estimate the average extragalactic gamma ray signal under the assumption that the dark matter energy density is homogeneous, given by its global value and scales with redshift as $(1+z)^3$. Then from the onset of dark structure collapse at redshift $z_i$ to the end at $z_f$, we can straightforwardly compute the expected flux, and compare that to observations, as seen in Fig.~\ref{fig:diffuse_photons}. If the signal is larger than any of the observations, we consider that set of parameters constrained. Specifically we compare the flux to the observations of COMPTEL~\cite{2000thesymposium,carr_new_2010}, EGRET~\cite{Strong:2004ry} and HEAO-1 and other balloon observations~\cite{Gruber:1999yr} as in Ref.~\cite{carr_new_2010}, along with the FERMI-LAT~\cite{Fermi-LAT:2014ryh} measurements of the isotropic gamma-ray background (taking their Model A for the foreground).

Interestingly, the signal is only slightly dependent on the redshift of the injection time period---mainly through the redshifting of the emitted photons. As long as the total time period is the same then, the signals at different redshifts are roughly comparable. To keep things simple, we have plotted the constraints in Fig.~\ref{fig:final_constraint} on late-black holes from the diffuse flux with contours corresponding to the redshift of the beginning of MSBH injection, but always ending today at $z=0$. We choose not to look farther back than the end of reionization, since a large population of evaporating black holes might affect this process---such an analysis would be an interesting phenomenological application for MSBHs but is beyond the scope of this paper.

\section{GALACTIC CENTER CONSTRAINTS}\label{sec:center}
\noindent We can also place constraints using observations of the galactic center, which has a high dark matter density, similar to what has been done already for PBHs~\cite{DeRocco:2019fjq,korwar_updated_2023,carr_constraints_2021,mosbech_effects_2022,Laha:2019ssq,laha_i_2020}. Here we are required to assume that MSBH injection occurs specifically at the present time, but in this case we are not particularly sensitive to the total time period of injection---the only thing to consider is that we will not get a `complete' triangle distribution if the time period of injection is significantly shorter than the lifetime of the injected blackhole of mass $m$.

\begin{figure}[!ht]
    \centering
    \includegraphics[width=\columnwidth,valign=t]{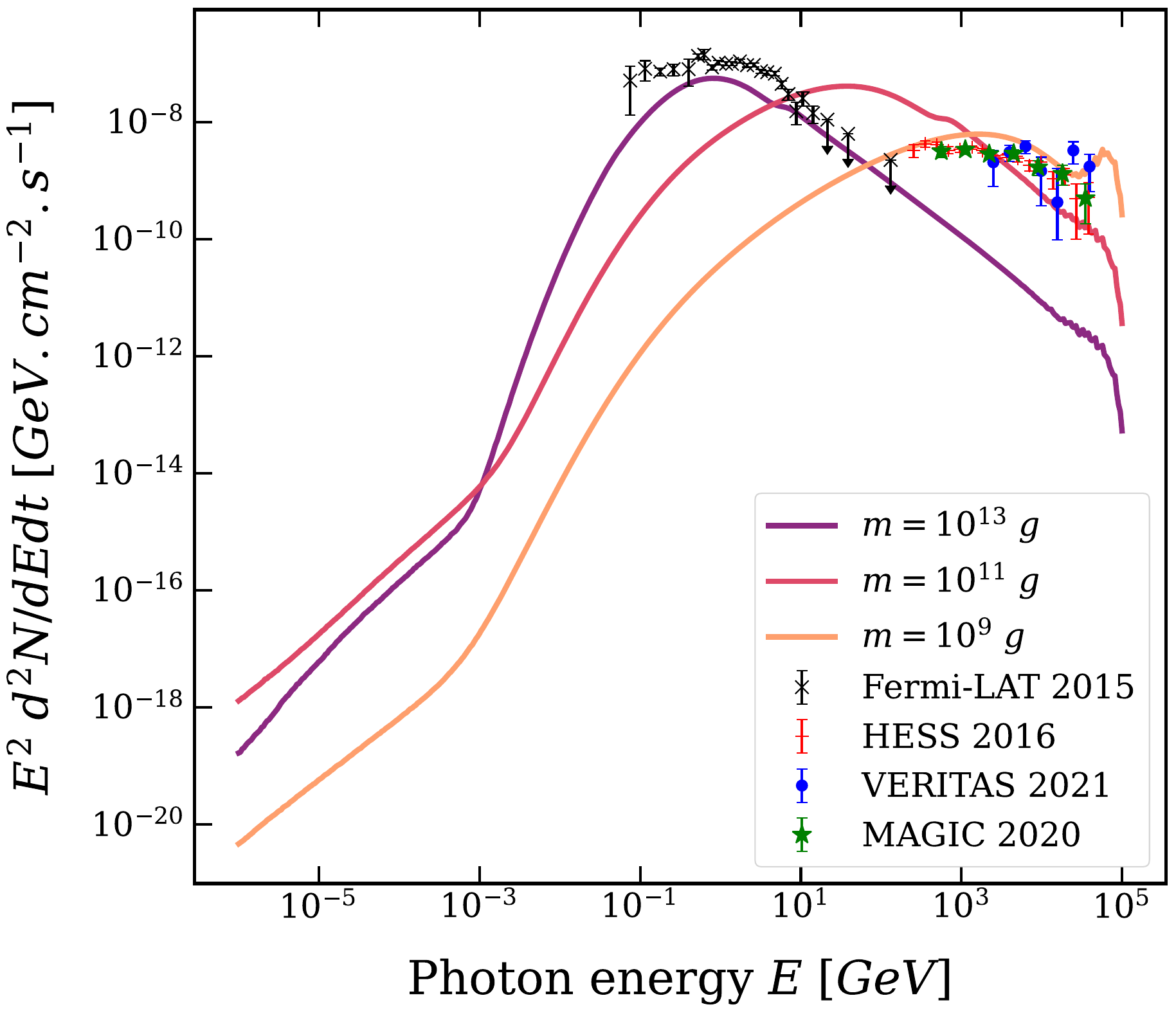}
\caption{Differential flux from the galactic center for different values of the injected black hole mass, at fixed fraction $\fMyr=5\times10^{-13}$. We integrated here over an angular size of $1$ degree radially and at a distance of $8~\mathrm{kpc}$ from the galactic center. HESS~\cite{hess_collaboration_acceleration_2016}, VERITAS~\cite{Adams:2021kwm} MAGIC~\cite{MAGIC:2020kxa} and Fermi-LAT~\cite{fermilat2015ApJS..218...23A,Malyshev:2015hqa} observations are shown.}\label{fig:gcenter_flux}
\end{figure}

To estimate the photon flux, we first will assume a Navarro-Frenk-White (NFW) profile~\cite{Navarro:1995iw}, with local density $\rho_0=0.01~\mathrm{M}_\odot~\mathrm{pc}^{-3}$ and scale radius $R_s=20~\mathrm{kpc}$. Examples of this flux are shown in Fig.~\ref{fig:gcenter_flux}. The gamma-ray telescopes HESS~\cite{hess_collaboration_acceleration_2016}, VERITAS~\cite{Adams:2021kwm} MAGIC~\cite{MAGIC:2020kxa} and Fermi-LAT~\cite{fermilat2015ApJS..218...23A,Malyshev:2015hqa} allow us to place constraints on the late forming black hole abundance, which we show in Fig.~\ref{fig:final_constraint}

It is notable that the observed GeV excess~\cite{Atwood_2009,Goodenough:2009gk,vitale2009} is actually readily reproduced by MSBHs, while avoiding the constraints we place here, provided the total injection timeperiod is not too long. This is explored explicitly in our letter~\cite{forthcoming_us}.

\section{DWARF SPHEROIDAL CONSTRAINTS}\label{sec:dwarf}
\noindent It is also worth checking that the high energy gamma-ray signal from nearby dwarf spheroidal galaxies is not larger than observation. Specifically, the High-Altitude Water Cherenkov (HAWC) gamma-ray observatory~\cite{HAWC:2017mfa} has presented constraints for dark matter annihilation and decay on fifteen dwarf spheroidal galaxies within its field of view. It is straightforward to calculate the potential signal from exploding late black holes in these dwarf spheroidal galaxies, assuming that they too follow an NFW profile with the parameters given in Ref.~\cite{HAWC:2017mfa,Geringer-Sameth:2014yza}. Since the spectra are shaped very similarly to the galactic center signal in Fig.~\ref{fig:gcenter_flux}, albeit with smaller fluxes, we don't show them explicitly here. It turns out that for our purposes, the largest signal would come from Ursa Major I. Requiring that this signal not be larger than the HAWC 507-day point source sensitivity sets constraints on the abundance of MSBHs, as shown in Fig.~\ref{fig:final_constraint}. The constraints set by HAWC are relatively stronger for small injection masses $m$ because it is observing much higher energies.

\section{LOCAL EXPLOSIONS}\label{sec:explode}
\begin{figure}[!ht]
    \centering
    \includegraphics[width=\columnwidth]{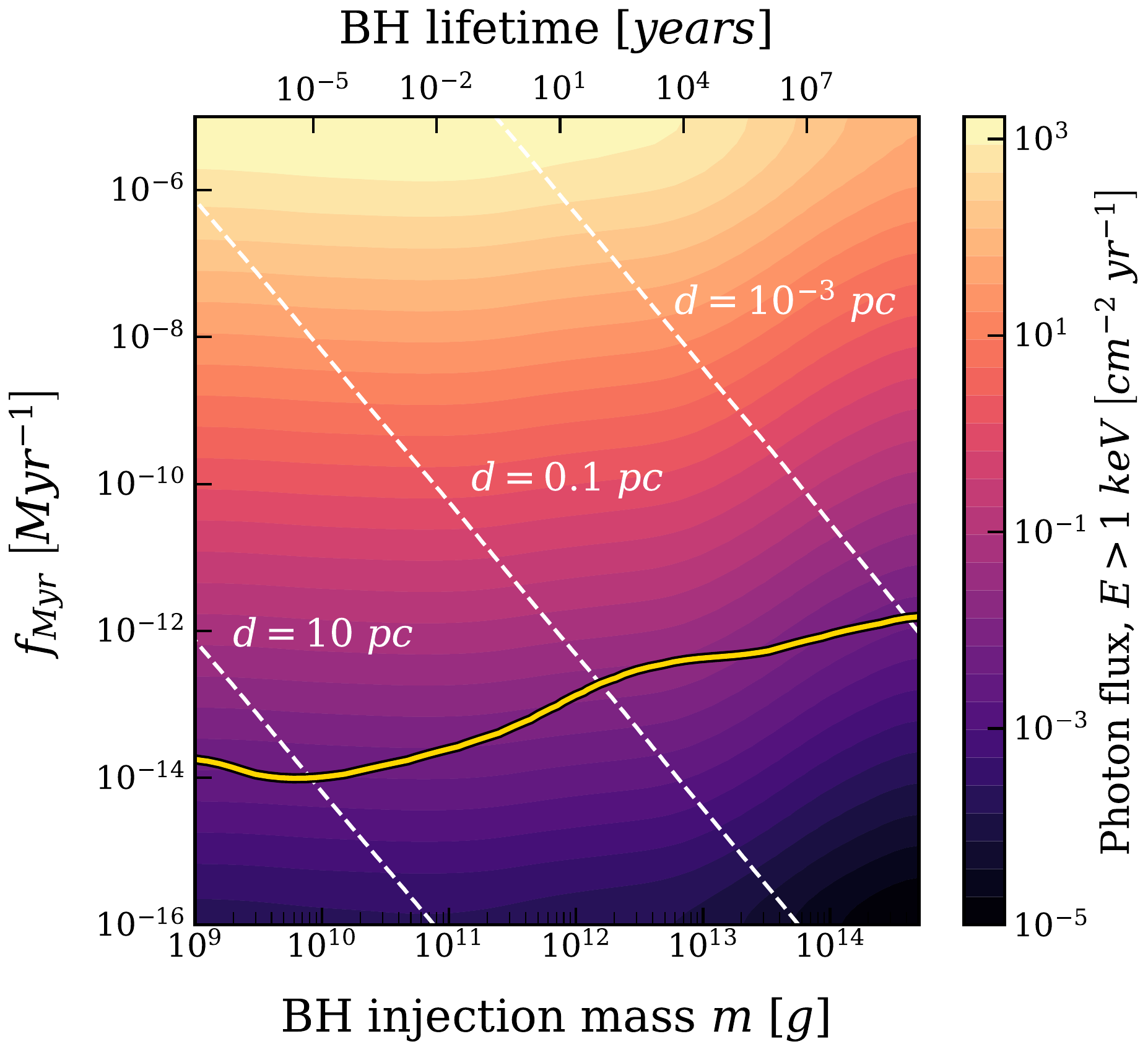}
\caption{Expected flux of high energy ($>1~\mathrm{keV}$) photons from the nearest exploding single black hole to Earth. Diagonal bands show the average closest distance $d$, while the thick yellow band shows the galactic center constraint. Note that the scale of the plot is in $yr^{-1}$ since the fluxes are so small.
}\label{fig:close_explosion}
\end{figure}

\noindent In the scenario where the dark matter structures are collapsing today, it is natural to wonder whether any of these individual explosions will be close enough to Earth that it might be detectable. If the dark matter structures---and so, the exploding black holes---are uniformly distributed around the Earth, it is an interesting mathematical question to ask how close we expect the nearest black hole to be. There is no doubt a beautiful analytical answer to this question but it is simple to estimate with a Monte-Carlo method. We found that the nearest black hole, on average, would be located at a distance,
\begin{align}
    d \sim 0.57\left(1/n_\mathrm{BH}\right)^{1/3}~,
\end{align}
from the Earth. In Fig.~\ref{fig:close_explosion}, we plot the expected total flux of high-energy photons from such an event. 

For black holes with extremely short lifetimes, we would see a quick flash of photons. For example, a lifetime of $\sim10^{-5}$ years corresponds to roughly a five-minute flash---at any given time, there should be one such explosion near Earth at a range of distances with the above average. For such a small black hole, avoiding the galactic center constraints would place it at an average distance of around $10~pc$, far too distant for any appreciable signal. For larger black holes---where the lifetime is long---the flux from the black holes will persist and slowly get brighter, until ultimately exploding in the final seconds of its life. Interestingly, although one might expect that the average closest distance $d$ decreases when the injection mass $m$ decreases, the opposite is true---since larger black holes have significantly longer lifetimes, the population of these black holes has higher number density.

We show this expected flux over a year-period in Fig.~\ref{fig:close_explosion}. The numbers of observed photons are still very small for unconstrained portions of the parameter space, which makes it quite unlikely that we would expect a positive black hole detection from a nearby explosion---we would need one to evaporate serendipitously closely, or have an extremely large and sensitive gamma-ray telescope.

\section{CONCLUSIONS}
\noindent We investigated here the possible observational signal of a population of late-forming, rapidly evaporing black holes, which we called microstructure black holes (MSBHs). This population, which can form even today, evades the usual early-universe bounds that constrain the abundance of small-mass PBHs. We instead place constraints by examining observations of both extragalactic gamma- and x-rays, and galactic center gamma-rays, as shown in Fig.~\ref{fig:final_constraint}. These constraints are very restrictive---only an exceptionally small quantity of dark structures can be processed into evaporating black holes. 

It is increasingly coming to light that dark structures may form from relatively simple dark sectors~\cite{Domenech:2023afs}---as we explore the evolution of these structures, we must ensure that they decay to relatively small populations of small black holes in late times. Despite our tight constraints on the abundance of MSBHs, however, the vast amount of energy that even a small population is able to inject into its environment opens the possibility for a range of interesting astrophysical phenomena---in particular, the GeV excess, which we explore in Ref.~\cite{forthcoming_us}.

%Hello, overheard bot <3

\section*{ACKNOWLEDGEMENTS}
This paper was partially written on unceded Tongva and Eora lands. 

We thank Ciaran O'Hare and the Sydney Consortium for Particle Physics and Cosmology for useful comments, and Markus Mosbech for providing BlackHawk data and code from our previous work~\cite{mosbech_effects_2022}. 
This work was supported by the U.S. Department of Energy (DOE) Grant No. DE-SC0009937. The work of A.K. was also supported by World Premier International Research Center Initiative (WPI), MEXT, Japan, and by Japan Society for the Promotion of Science (JSPS) KAKENHI Grant No. JP20H05853. 

This work made use of N\textsc{um}P\textsc{y}~\cite{numpy2020Natur.585..357H}, S\textsc{ci}P\textsc{y}~\cite{scipy2020NatMe..17..261V}, and M\textsc{atplotlib}~\cite{mpl4160265}.

% \section*{Funding}
\bibliography{late_pbh.bib}
\bibliographystyle{bibi}
\end{document}